\documentclass[final]{siamltex}
\usepackage{amsmath}
\newcommand\ignore[1]{{}}

\newcommand{\from}{\leftarrow}

\newcommand{\comment}[1]{}

\DeclareMathOperator{\LL}{L}
\DeclareMathOperator{\Pos}{Pos}
\DeclareMathOperator{\Posv}{Pos_v}
\DeclareMathOperator{\Posnv}{Pos_{nv}}
\DeclareMathOperator{\StructDiff}{StructDiff}
\DeclareMathOperator{\height}{height}
\DeclareMathOperator{\Vars}{Vars}
\DeclareMathOperator{\Sol}{Sol}

\DeclareMathOperator{\selPos}{selPos}
\DeclareMathOperator{\Min}{min}
\DeclareMathOperator{\Max}{max}
\DeclareMathOperator{\Prefixes}{Prefixes}
\newcommand{\OO}{\ensuremath{\mathcal{O}}}

\title{Deciding Regularity of the Set of Instances of a Set of Terms with
Regular Constraints is EXPTIME-Complete}

\author{Omer Gim\'enez\thanks{%
Universitat Polit{\`e}cnica de Catalunya,
Barcelona, Spain
({\tt $\{$ogimenez,ggodoy$\}$@lsi.upc.edu}).
The second author was supported by
Spanish Min.\ of Educ.\ and Science by the
FORMALISM project (TIN2007-66523) and by the
LOGICTOOLS-2 project (TIN2007-68093-C02-01).
}
\and Guillem Godoy\footnotemark[1] \and
Sebastian Maneth\thanks{
NICTA and University of New South Wales, 
Sydney, Australia ({\tt sebastian.maneth@nicta.com.au})}
}

\begin{document}

\maketitle

\begin{abstract}
Finite-state tree automata are a well studied formalism for
representing term languages. This paper studies the problem of
determining the regularity of the set of instances of a finite set of
terms with variables, where each variable is restricted to
instantiations of a regular set given by a tree automaton. The problem
was recently proved decidable, but with an unknown
complexity. Here, the exact complexity of the problem is determined
by proving EXPTIME-completeness. 
The main contribution is a
new, exponential time algorithm that performs various exponential
transformations on the involved terms and tree automata, and decides
regularity by analyzing formulas over inequality and height
predicates.
\end{abstract}



\pagestyle{myheadings}
\thispagestyle{plain}
\markboth{O. GIMENEZ, G. GODOY AND S. MANETH}{REGULARITY OF
TERMS WITH REGULAR CONSTRAINTS}  

\section{Introduction}
Finite representations of infinite sets of terms are 
useful in many areas of computer science. 
The choice of formalism for this purpose 
depends on its expressiveness, 
but also on its computational properties. 
Finite-state tree automata (TA)~\cite{gecste97,tata2007}
are a well studied formalism for representing term languages,
due to their good computational and expressiveness properties.
They characterize the ``regular term languages'',
a classical concept used, e.g., 
to describe the parse trees of a context-free grammar 
or the well-formed terms over a sorted signature~\cite{mezwri67}, 
to characterize the solutions of formulas in monadic 
second-order logic~\cite{don70}, and 
to naturally capture type formalisms for tree-structured XML
data~\cite{DBLP:journals/toit/MurataLMK05,bencashospievan08}.
Similar to the case of regular sets of words, 
regular term languages have numerous convenient properties
such as closure under Boolean operations (intersection,
union, negation), decidable properties such as finiteness and
inclusion, and they are characterized by many different formalisms
such as regular grammars, 
regular term expressions, 
congruence classes of finite index,
deterministic bottom-up TA,
nondeterministic top-down TA,
or sentences of monadic second-order logic~\cite{tata2007}. 
Deterministic TA, for instance, can be effectively
minimized and give rise to efficient parsing.

When the used formalism for representing an infinite set of terms
is not a TA, it is often expedient to decide whether
the represented set is in fact regular. 
A simple and natural way of describing
an infinite set of terms, is through the use of ``patterns''. 
A pattern is a term with variables; it describes 
all terms obtained by replacing the variables by (variable-free)
terms; see, e.g.,~\cite{lasmar87,kucrus99}, and the references
given there.
Term patterns are used for pattern matching in
most modern programming languages, and were already
present in very early languages such as LISP.
They are a central concept in compiling, natural language processing,
automated deduction, term rewriting, etc. In some of these
applications, variables in patterns are restricted to be replaced
by terms in a regular language. E.g. in a programming language with
regular types (see, for
instance,~\cite{DBLP:journals/jfp/HosoyaP03,DBLP:journals/toplas/HosoyaVP05}),
variable instances might be constrained to regular 
term languages. Typically, term patterns in a 
programming language must be linear (i.e.,
every variable occurs at most once) in order to 
guarantee that the resulting type is regular. Our result shows
that even if non-linear patterns are allowed (which is the 
case in logic programming languages such as Prolog), one can
statically determine regularity, i.e., the existence of
an exact regular type, in exponential time.

More precisely,
we consider the problem of determining the regularity
of the set of instances of a set of terms with regular constraints,
which we abbreviate as the ``RITRC'' problem. A particular case of
this problem, in which variables can be replaced by \emph{arbitrary}
terms (without variables), was considered in~\cite{lasmar87} and shown
to be coNP-complete (cf. also~\cite{kucrus99}). The general
RITRC problem was recently proved decidable~\cite{godmantis08}.
The complexity of their decision procedure was
left open in~\cite{godmantis08}, but can easily be
seen to exceed exponential time. 
Moreover, their solution is based on a rather general result
of~\cite{ComonDelor94} 
about first-order formulas with regular constraints,
for which the complexity is not known.

In this paper, we determine the complexity of the RITRC problem
by proving that it is EXPTIME-complete. At the beginning of
Section~\ref{sect:main} we show that the RITRC problem is
EXPTIME-hard. 
This is done via a straightforward reduction from the
finite intersection emptiness problem for tree automata. 
The remaining part of Section~\ref{sect:main} describes an
EXPTIME algorithm solving the problem, starting with an overview of
it in Section~\ref{subsect:overview}.
In summary, the
algorithm first changes the regular constraints from several TA to one
single tree automaton (of exponential size) with special properties.  
It then picks a non-linear term $s$ from the given set $S$ of
terms, and checks the ``infinite instances property of $s$ in $S$'':
are there infinitely many instantiations of a non-linear
variable $x$ in $s$, which are \emph{not} instances of
$S-\{s\}$ (under the regular constraints)? 
If the infinite instances property holds for some $s$ in $S$, then
our algorithm stops and we know that the set of terms 
represented by $S$ (under the regular constraints) is \emph{not
  regular}.
Otherwise, we can replace $s$ by a new term $s'$ that is
linear in the variables, i.e., which does not contain
duplicated variables. Roughly speaking, 
our algorithm then starts over again,
with the new set $(S-\{s\})\cup\{s'\}$.
In this way, the algorithm will construct a set $S'$
of terms in which all terms
are linear in the variables, if and only if the represented
set is \emph{regular}.
To check the infinite instances property of $s$ in $S$, we
instantiate the term $s$ at all non-variable positions of terms in 
$S-\{s\}$, and then formulate inequality constraints of
the resulting terms with terms of $S-\{s\}$. 
It is a non-trivial task to efficiently solve such inequality
constraints.
In fact, 
in order to solve systems of such inequality constraints in EXPTIME,
it was a crucial step for us to introduce additional height constraints
on the variables of the inequality constraints. The final formula
$F$ over height and inequality predicates 
characterizes all instances of $s$ that are
not instances of terms in $S-\{s\}$. 
Our algorithm solves the RITRC problem
in exponential time by iteratively constructing and solving such
formulas $F$.

\section{Preliminaries}

The size of a set $S$ is denoted by $|S|$.
A \emph{signature} consists of an alphabet $\Sigma$, 
i.e., a finite set of symbols, together with a mapping that assigns to each
symbol in $\Sigma$ a natural number, its \emph{arity}.
We write $\Sigma^{(k)}$ to denote the subset of symbols in
$\Sigma$ that are of arity $k$, and we
write $f^{(k)}$ to denote that $f$ is a symbol
of arity $k$. 
The \emph{set of all terms over $\Sigma$} is denoted $T_\Sigma$ and is
inductively defined as the smallest set $T$ such that for
every $f\in\Sigma^{(k)}$, $k\geq 0$, and $t_1,\dots,t_k\in T$,
the term $f(t_1,\dots,t_k)$ is in $T$. For a term of the form
$a()$ we simply write $a$.
For instance, if $\Sigma=\{f^{(2)},a^{(0)}\}$ then 
$T_\Sigma$ is the set of all terms that represent binary trees
with internal nodes labeled $f$ and leaves labeled $a$.
We fix the set $X=\{x_1,x_2,\dots\}$ of variables, i.e., 
any set $V$ of variables is always assumed to be a subset of $X$.
The set of terms over
$\Sigma$ with variables in $X$, denoted $T_\Sigma(X)$, is the
set of terms over $\Sigma\cup X$ where every symbol in $X$ has
arity zero.
By $\Vars(s)$ we denote the set of variables that occur in $s$.
By $|t|$ we denote the size of $t$, defined recursively as
$|f(t_1,\ldots,t_k)|=1+|t_1|+\ldots+|t_k|$ for
each $f\in\Sigma^{(k)}$, $k\geq 0$ and $t_1,\dots,t_k\in T_\Sigma$,
and $|x|=1$ for each $x$ in $X$.
By $\height(t)$ we denote the height of $t$, defined recursively as
$\height(f(t_1,\ldots,t_k))=1+\Max(\height(t_1),\ldots,
\height(t_k))$ for each
$f\in\Sigma^{(k)}$, $k\geq 1$ and $t_1,\dots,t_k\in T_\Sigma$,
$\height(a)=0$ for each $a\in\Sigma^{(0)}$, and
$\height(x)=0$ for each $x\in X$.
Given a term $f(t_1,\dots,t_k)\in T_\Sigma$, 
its set of positions $\Pos(t)$ equals
$\{\varepsilon\}\cup_{1\leq i\leq k}\{i.p\mid p\in \Pos(t_i)\}$.
Here, $\varepsilon$ denotes the root node, and
$p.i$ denotes the $i$th child of position $p$.
The subterm of $t$ at position $p$ is denoted by
$t/p$, and the symbol of $t$ at position $p$ is denoted
by $t[p]$; we say that $p$ is labeled by $t[p]$.
For instance, for $s=g(f(a,b),c)$, $s/1$ equals $f(a,b)$
and position $1.2$ is labeled by $b$.
For a set $\Gamma$, we use
$\Pos_\Gamma(t)$ to denote the set of positions of $t$ that
are labeled by symbols in $\Gamma$.
In particular, we define for $t\in T_\Sigma(X)$
the sets $\Posv(t)$ and
$\Posnv(t)$ of variable positions
and non-variables positions as 
$\Pos_X(t)$ and 
$\Pos(t)-\Pos_X(t)$, respectively.
E.g., for $s$ as above, $\Pos_{\{c\}}(s)=\{2\}$ and
$\Posv(s)=\emptyset$.
When a position $p$ is of the form $p_1.p_2$,
we say that $p_1$ is a prefix of $p$.
For a set of positions $P$, we denote by
$\Prefixes(P)$ the set $\{p~\mid~\exists p': p.p'\in P\}$.
For terms $s,t$ and $p\in\Pos(s)$, we denote by
$s[p\leftarrow t]$ the result of replacing the subterm
at position $p$ in $s$ by the term $t$.
For instance, $f(f(a,a),a)[1\leftarrow a]=f(a,a)$.

A (\emph{deterministic})
\emph{tree automaton}
(\emph{over $\Sigma$}), DTA for short, is a tuple 
$A=\langle Q,F,\Sigma,\delta\rangle$ 
where $Q$ is a finite set of states,
$F\subseteq Q$ is the set of accepting states, $\Sigma$ is a signature,
and $\delta$ is a set of transitions of the form
$f(q_1,\dots,q_k)\to q$, where
$f\in\Sigma^{(k)}$, $k\geq 0$, and $q,q_1,\dots,q_k\in Q$.
Moreover, for each $f\in\Sigma$ and each $q_1,\ldots,q_k\in Q$
there exists at most one (and at least one if the automaton
is \emph{complete}) $q$ such that
$f(q_1,\ldots,q_k)\to q$ is in $\delta$.
The language $\LL(A)$ recognized by $A$ is the set
$\{t\in T_\Sigma\mid A(t)\in F\}$ where $A(t)$ is 
recursively defined as
$A(f(t_1,\dots,t_k))=q$ if 
$f\in\Sigma^{(k)}$, $k\geq 0$, $t_1,\dots,t_k\in T_\Sigma$,
$f(q_1,\dots,q_k)\to q$ is a transition in $\delta$, and,
for each $i\in\{1,\dots,k\}$, $q_i=A(t_i)$.
Note that, when $A$ is not complete, $A(t)$ might be undefined.
We also define, for $q\in Q$, the set
$\LL(A,q)=\{t\in T_\Sigma\mid A(t)=q\}$ of terms for which $A$
arrives to state $q$.
Note that $\LL(A,q)\cap \LL(A,q')=\emptyset$ for all $q\not= q'$.
We also extend $A(t)$ to terms $t$ in
$T_{\Sigma\cup Q}$ by assuming that the states $q\in Q$
have arity $0$ and $A(q)=q$ for each $q\in Q$.
A set of terms $L\subseteq T_\Sigma$ is \emph{regular}
if there exists a DTA $A$ such that $L=\LL(A)$.
The size $|\tau|$ of a transition $\tau=(f(q_1,\dots,q_k)\to q)$ is
$k+2$ and 
the size $|A|$ of $A$ is
$|Q|+\sum_{\tau\in\delta}|\tau|$.

Given a DTA, it is decidable whether its recognized language is (i)
empty, (ii) finite, or (iii) has cardinality $k$, for a given $k$. The
corresponding constructions all run in polynomial time and are
straightforward generalizations of the ones for classical finite
(word) automata; proofs can be found in Theorems 1.7.4, 1.7.6, and
1.7.10 of~\cite{tata2007}. The following computational problems,
together with the running times, are a consequence of the same proofs.

\begin{lemma}
\label{lem:DTA_folklore}
Let $A=\langle Q,F,\Sigma,\delta\rangle$ 
be a DTA and $k$ a natural number.
Each of the following sets can be computed in polynomial time:
$\operatorname{non-emptyStates}(A):=\{q\in Q\mid \LL(A,q)\not=\emptyset\}$
in $\OO(|Q|+|\delta|)$,
$\operatorname{infiniteStates}(A):=\{q\in Q\mid |\LL(A,q)|=\infty\}$
in $\OO(|Q|\cdot|\delta|)$, and
$\operatorname{countUpto}(A,k):=\{(q,min(|\LL(A,q)|,k))\mid q\in Q\}$
in $\OO(|Q|\cdot|\delta|)$.
\end{lemma}

{\bf Sets of Terms with Regular Constraints}\quad
Let $V\subseteq X$ be a finite set 
of variables and $\Sigma$ a signature.
A \emph{regular constraint} (\emph{over $V$ and $\Sigma$})
is a mapping $M$ that associates to every $x\in V$ a
DTA over $\Sigma$.
A \emph{solution of $M$} is a mapping
$\varphi: V\to T_\Sigma$ such that, for each $x\in V$,
$\varphi(x)\in \LL(M(x))$.
A \emph{set of terms with regular constraints} 
(\emph{over $V$ and $\Sigma$}) is a pair
$\langle S,M\rangle$ where $S$ is a finite subset of 
$T_\Sigma(V)$ and $M$ is a regular constraint 
over $V$ and $\Sigma$.
The language $\LL(\langle S,M\rangle)$ 
of $\langle S,M\rangle$ is defined as
$\{t\mid\exists \varphi,s:(
t=\varphi(s)\wedge 
s\in S \wedge
\varphi\mbox{ is a solution of }M
)\}$.
A term in $\LL(\langle S,M\rangle)$ is also called
an \emph{instance} of $\langle S,M\rangle$.

The following result is due to~\cite{lasmar87},
cf. also~\cite{kucrus99}.

\begin{proposition}\label{prop:trivial_constraint}
Let $V\subseteq X$, $S$ a finite subset of $T_\Sigma(V)$, and
$M$ the regular constraint that maps every $x\in V$ to
the trivial DTA that recognizes $T_\Sigma$.
Regularity of $\LL(\langle S,M\rangle)$ is coNP-complete.
\end{proposition}

When analyzing complexity, with $\|S\|$ we refer to the
sum of sizes of all terms in $S$, and with $\|M\|$ we
refer to the sum of sizes of all DTA in the image
of $M$. With $|S|$ and $|M|$ we refer, as usual, to
the number of elements in the sets $S$ and $M$ (i.e.\
number of pairs of the set defining the mapping $M$).
We also do the following assumption in order to ease
the complexity analysis.

{\bf Assumption:} The maximum arity of a function symbol
in $\Sigma$ is $2$. It is well known that any arbitrary tree
can be coded as a binary tree of essentially the same size.
Usual such codings (such as the one taking first-child to 
left-child and next-sibling to right-child) preserve 
regularity of sets of terms (see, e.g., Section~8.3.1 in
~\cite{tata2007}); moreover, 
it can be seen easily that the transformation 
of the regular constraints into this
new binary signature produces an at most quadratic size
increase.

\section{Regularity of the instances of a set
of terms with regular constraints}
\label{sect:main}

Let $\langle S, M\rangle$ be a set of terms with
regular constraints.
The ``\emph{regularity of the instances of a set of terms with
regular constraints problem}'', RITRC for short, asks 
whether or not the set $\LL(\langle S, M\rangle)$ is regular.
We know, by Proposition~\ref{prop:trivial_constraint},
that RITRC is coNP-complete in the particular case that $M$ maps each
variable to a DTA that accepts all terms.
In general, i.e., \emph{with} regular constraints, 
decidability of RITRC was proved in~\cite{godmantis08}; however, the complexity
remained open. The algorithm of~\cite{godmantis08} does not run in
exponential time, and in fact it has a far worse complexity.
In this section we show
that RITRC is EXPTIME-complete.
We start with the easy part by showing that RITRC is EXPTIME-hard.

\begin{theorem}
RITRC is EXPTIME-hard.
\end{theorem}
\begin{proof}
Let $\Sigma$ be a signature with
$\Sigma^{(2)}\not=\emptyset$ and let $A_1,\dots,A_n$ be DTAs
over $\Sigma$. 
It is well known that testing whether
$\LL(A_1)\cap\dots\cap \LL(A_n)=\emptyset$
is EXPTIME-complete, cf.~Theorem~1.7.5 of~\cite{tata2007}.
It follows that
``universality of union'', i.e., testing whether
$\LL(A_1)\cup\dots\cup \LL(A_n)=T_\Sigma$
is EXPTIME-complete. This is because a DTA can
easily be complemented in polynomial time (first
complete the DTA by adding, for
any missing transition, a transition to a new 
``sink'' state; second, change $F$ into
$Q-F$). 
We now reduce universality of union to RITRC.
Let $A$ be any fixed DTA that
recognizes $T_\Sigma$ and let $f\in\Sigma^{(2)}$.
The set of terms with regular constraints $\langle S, M\rangle$,
where 
\begin{align*}
S &= \{f(f(x,x),y),f(x_1',x_1),\ldots,f(x_n',x_n)\}\\
M &= \{x_1\mapsto A_1,\ldots,x_n\mapsto A_n,
x\mapsto A, y\mapsto A,\\ 
&\ \qquad\qquad
x_1'\mapsto A,\ldots,x_n'\mapsto A\},
\end{align*}
is regular if and only if
$\cup_{1\leq i\leq n}\LL(A_i)=T_\Sigma$.
To see this, consider first the case where 
$\cup_{1\leq i\leq n}\LL(A_i)=T_\Sigma$.
Then $\LL(\langle S,M\rangle)=
\LL(\langle \{f(x_1',x_1),\ldots,f(x_n',x_n)\},M\rangle)=
\{f(s,t)\mid s,t\in T_\Sigma\}$, which is regular. 
In the other case, let $t$ be in
$T_\Sigma - \cup_{1\leq i\leq n}\LL(A_i)$.
Intersect $\LL(\langle S,M\rangle)$ with
the regular set 
$\{f(s,t)\mid s\in T_\Sigma\}$.
Since regular term languages are closed under intersection,
the resulting set would be regular, if $\LL(\langle S,M\rangle)$ was;
but, the resulting intersection is
$\{f(f(t',t'),t)\mid t'\in T_\Sigma\}$.
By standard pumping arguments 
(see, e.g., Example~1.2.1 of~\cite{tata2007})
this set is not regular. Thus, $\LL(\langle S, M\rangle)$ is not
regular in this case.
\end{proof}

Proving that RITRC is in EXPTIME is considerably more
complicated.

\subsection{Overview of our algorithm for RITRC}
\label{subsect:overview}

\textbf{Algorithm in \cite{godmantis08}.} In \cite{godmantis08}
decidability of RITRC was proved.
We first explain the idea of that proof, and why it
does not give rise to an EXPTIME algorithm.
Then we give an overview of
the algorithm presented in this paper.
The following is the basic property used
for deciding RITRC in~\cite{godmantis08} (and here).

\begin{definition}\label{def:infinst}
Let $\langle S,M\rangle$ be a set of terms with regular constraints.
The term $s\in S$ satisfies
the {\em infinite-instances property} in $\langle S,M\rangle$
if some variable $x$ has multiple occurrences in $s$, and
there exists infinitely many instances $\varphi_1(s),\varphi_2(s),\ldots$
of $\langle \{s\},M\rangle$ which are not instances of
$\langle S-\{s\},M\rangle$ and all of them different on $x$,
i.e., $\varphi_i(x)\not=\varphi_j(x)$ for all $i\not=j$.
\end{definition}

In~\cite{godmantis08} it was shown that 
the infinite-instances property is decidable
and that it implies non-regularity of $\langle S,M\rangle$.
To decide RITRC, the algorithm of~\cite{godmantis08} first
looks for a term in $S$ with multiple occurrences of some variable
$x$ satisfying $|\LL(M(x))|=\infty$.
If no such term exists, then it stops concluding
regularity of $\LL(\langle S,M\rangle)$ (note that
in this case
$\LL(\langle \{s\},M\rangle)$ is regular for each term $s$ in $S$, and
regular sets are closed under union). Otherwise, 
it checks the infinite-instances property of $s$
in $\langle S,M\rangle$.
In the affirmative case, it stops concluding
non-regularity of $\LL(\langle S,M\rangle)$.
In the negative case, there are only a finite number of
possible instantiations $\varphi(x)$ of each duplicated variable $x$
in $s$ providing a term in
$\LL(\langle \{s\},M\rangle)$ and not in
$\LL(\langle S-\{s\},M\rangle)$. Thus, by replacing $s$
by a finite number $\{s_1,\ldots,s_k\}$
of instantiations of $s$, the represented language
$\LL(\langle S,M\rangle)$ is preserved,
and we obtain less duplicated variables.
The algorithm in \cite{godmantis08} decides
regularity of $\LL(\langle S,M\rangle)$ by
iterating this process.

\textbf{Estimating the complexity.}
To determine the complexity of the previous algorithm, we
need to know how large is the number $k$ of instantiations of $s$, how
large the terms $s_1,\dots,s_k$ are, and, of course,
how expensive it is to decide the infinite instances property.
In~\cite{godmantis08}, the latter is
solved through a result of~\cite{ComonDelor94} 
about first-order formulas with regular constraints. 
The precise complexity of this result
of~\cite{ComonDelor94} is not known, but it is expected to be
higher than that of solving the infinite-instances property, since
it solves a more general problem.
We therefore devise our own algorithm for checking
this property. But, also the sum of sizes of the
terms $s_1,\dots,s_k$ poses a problem, as it can grow
iterated exponential, so the algorithm in \cite{godmantis08}
is certainly not in EXPTIME.
One of the ideas of our new algorithm is hence
\emph{not} to replace $s$ by $\{s_1,\dots,s_k\}$. 
Instead, we are able to find a ``small'' number $h$
(which depends on $S$ and $M$) such that all terms
$s_i$ are guaranteed to be of height smaller than $h$. 
To take advantage of this fact, we add a 
new kind of constraint to $\langle S,M\rangle$ which allows
duplicated variables $x$ of $s$ to be replaced only by ``small''
terms. The algorithm then continues on with this new system (called
\emph{restricted regular constraints}, see Definition~\ref{def:restregconstr}),
which has regular constraints plus height constraints on
the variables.

\textbf{Infinite-instances algorithm.}
How do we check the infinite-instances property of $s$ 
in $\langle S,M\rangle$?
In Sections~\ref{sec:determining-term},~\ref{sec:subsumed-terms},
and~\ref{sec:formula-instances} we give an algorithm that solves
this problem under several assumptions. To begin with, we
require that the term $s$ is \emph{determined}
(see Definition~\ref{def:determined} for the
precise notion)
in all the non-variable positions
of terms in $S$. We also assume that the regular constraint $R$
is given by a single DTA $A$ (instead of the multiple
ones in the image of $M$), and a mapping that associates
variables with states of $A$. Finally, we require this
DTA $A$ to satisfy the \emph{$1$-or-$|S|$} property
of Definition~\ref{def:oneorkautomaton}, which says
that for any state $q$ of $A$, the cardinality of $\LL(A,q)$ is
either $1$, or it is greater than or equal to $|S|$. 
The reason for these assumptions is as follows.
In order to decide the infinite-instances property,
we compute a formula $\mathcal{F}$ whose solutions
are the instances in $\LL(\langle \{s\},R\rangle)$ that are not in
$\LL(\langle S-\{s\},R\rangle)$. This formula is a disjunction
of conjunctions of inequalities, where each conjunction has at most
$|S|-1$ inequalities. After some transformations on $\mathcal{F}$ by means
of a system of inference rules,
the variables $x$ with an associated state $q_x$ of $A$
satisfying $|\LL(A,q_x)|=1$ disappear. Thanks to the $1$-or-$|S|$
property, the remaining variables in $\mathcal{F}$ have at least $|S|$
possible instantiations. This fact is used to show that, for any
surviving conjunction in $\mathcal{F}$, there is a variable instantiation
that makes true the at most $|S|-1$ inequalities it is composed
of, and variables with infinite language have infinite choices.
Hence, we obtain that $s$ satisfies the infinite-instances
property in $\langle S, R\rangle$ if the transformed formula
$\mathcal{F}$ is not empty. 

\textbf{Overview of the algorithm.}
We give an outline of the EXPTIME algorithm that solves
RITRC for a given instance $\langle S_1,M_1\rangle$.
First of all, we transform $\langle S_1,M_1\rangle$
into $\langle S_2,R_1\rangle$ by preserving the
represented language, where $R_1$ is a {\em single regular constraint} (Definition~\ref{def:singleregconstr}),
and $S_2$ is the adaptation of $S_1$ from $M_1$ to $R_1$. Intuitively,
$\langle S_2,R_1\rangle$ is the same problem stated with a single
$1$-or-$|S_1|$ DTA; the sizes of both $S_2$ and $R_1$ can be
exponential with respect to the sizes of $S_1$ and $M_1$. This
transformation is described in Section~\ref{subsec:singleautomaton}. 
The single regular constraint $R_1$ is then
converted to a \emph{restricted regular constraint} $R_2$,
the new type of constraint, which we introduce
in Section~\ref{subsec:heightconstraints}, that takes account of
height restrictions.

The algorithm then proceeds as follows. At each step it picks
a term $s$ of $S_2$ without height constraints,
and with multiple occurrences of some variable
$x$ satisfying $|\LL(A,C(x))|=\infty$.
If no term of this kind exists, then it stops concluding
regularity of $\LL(\langle S_2,R_2\rangle)$.
Otherwise, it chooses a term $s$ satisfying the above conditions,
and checks the infinite-instances property of $s$ with respect to
$\langle S_2, R_2\rangle$. To do so, the algorithm loops over all possible
partial instantiations $s_i$ of $s$ in the non-variable positions of
$S_2$, and for each $s_i$, it finds a subset $S_3\subseteq S_2$, with 
$|S_3|=|S_1|-1$, such that $s_i$ has the infinite-instances property for 
$S_2$ if and only if it has the property for $S_3$. The fact that
$|S_3|$ is small allows to check the infinite-instances property 
in exponential time.
In the affirmative case the algorithm stops concluding
non-regularity of $\LL(\langle S_2,R_2\rangle)$.
If no determination $s_i$ satisfies the infinite-instances property,
the restricted regular constraint
$R_2$ is modified so as to impose height constraints on the
variables of $s$ with multiple occurrences. 
Since the number of terms with duplicated variables
and without height constraints decreases,
the iteration of this process decides
regularity of $\LL(\langle S_1,M_1\rangle)$.
A careful analysis of all the steps involved will show that the time
complexity is exponential.

\subsection{Simplification to a single DTA}\label{subsec:singleautomaton}

Recall from the preliminaries that we assume $\Sigma$ to be a fixed
but arbitrary signature containing no symbol of arity greater than $2$.
We start with a set of terms with regular constraints
$\langle S_1,M_1\rangle$ over a finite set of variables $V$.
Recall that $S\subseteq T_\Sigma(V)$ is a finite set of terms
and $M$ is a function that maps each $x\in V$ to a
DTA over $\Sigma$.
We now adapt this definition to a setting with
only one single DTA $A$, and where variables in $V$ are now
mapped to states in $A$.
Moreover, we do not need accepting states anymore and
simply drop them from $A$'s 
definition (a ``DTA without accepting states'').

\begin{definition}\label{def:singleregconstr}
A {\em single regular constraint} (\emph{over $V$ and $\Sigma$})
is a pair $R=\langle A,C\rangle$, 
where $A=\langle Q,\Sigma,\delta\rangle$ is a complete DTA
without accepting states and
$C$ is a mapping $C:V\to Q$. 
The \emph{size} $\|R\|$ of $R$ is $|V|+\|A\|$.
A {\em solution} of $R$ is a mapping
$\varphi:V\to T_\Sigma$ such that, for each $x\in V$, it holds that
$\varphi(x)\in \LL(A,C(x))$.
A {\em set of terms with single regular constraints} (\emph{over $V$
  and $\Sigma$}) is a pair $\langle S,R\rangle$, 
where $S$ is a finite subset of $T_\Sigma(V)$ and 
$R=\langle A,C\rangle$ is a single regular constraint
over $V$ and $\Sigma$.
The language $\LL(\langle S, R\rangle)$ of $\langle S,R\rangle$ is
defined as 
$\{t\mid\exists \varphi,s:(t=\varphi(s)\wedge 
s\in S \wedge\varphi\mbox{ is a solution of }R
)\}$. A term in $\LL(\langle S,R\rangle)$ is also called
an \emph{instance} of $\langle S,R\rangle$.
\end{definition}

Transforming a set of terms with regular constraints
$\langle S_1,M_1=\{x_1\mapsto A_1,\ldots,x_n\mapsto A_n\}\rangle$
into a set of terms with single
regular constraints $\langle S_2,R_1\rangle$ satisfying
$\LL(\langle S_2,R_1\rangle)=\LL(\langle S_1,M_1\rangle)$
is rather easy by considering the product automaton
$A=A_1\times\cdots\times A_n$.
But the size of $\langle S_2,R_1\rangle$ can be 
exponential in the size of $\langle S_1,M_1\rangle$. 
Moreover, it follows from Proposition~\ref{prop:trivial_constraint}
that regularity of $\LL(\langle S_2,R_1\rangle)$ is at least NP-hard.
Hence, it is not enough to have an EXPSPACE-reduction from one problem
to the other if we want to obtain an EXPTIME algorithm for the initial
problem. 

Thus, in the translation from $\langle S_1,M_1\rangle$
into $\langle S_2,R_1\rangle$ we keep in mind some additional
properties obtained by the transformation process.
For instance, the terms in $S_2$ are very similar to those
in $S_1$ because they are obtained through variable renamings; we
call this ``structural similarity''. Moreover, as mentioned 
in the outline of Section~\ref{subsect:overview}, we want the
DTA $A$ to have the ``$1$-or-$n$'' property, with $n=|S_1|$.
We proceed to define both properties.

\begin{definition} \label{def:structsim}
Let $V,V'$ be sets of variables. 
A total function $\rho:V\to V'$ is a \emph{variable renaming}
if it is injective, i.e., $\rho(x)\not=\rho(y)$ for $x\not=y$.
For a term $s$, $\rho(s)$ is the term obtained from $s$ by
replacing in $s$ each variable $x\in V$ by $\rho(x)$.
Two terms $s$ and $t$ are \emph{structurally similar},
denoted by $s=_\Sigma t$, if $t=\rho(s)$ for a variable renaming
$\rho$.
For a set of terms $S$, $\StructDiff(S)$ is the
maximum number of non-structurally similar terms in $S$, i.e.,
$\StructDiff(S)=
{\tt max}_{S'\subseteq S\wedge(s,t\in S'\Rightarrow s\not=_\Sigma
  t)}|S'|$.
Given a single regular constraint $R=\langle A,C\rangle$ we say
that two terms $s$ and $t$ are \emph{structurally equal} 
(\emph{with respect to $R$}) if
they are structurally similar, and
$C(s[p])=C(t[p])$ for all $p\in\Posv(s)$.
\end{definition}

Note that if $s$ and $t$ are structurally equal with
respect to $R$, then $\LL(\langle \{s\},R\rangle)=\LL(\langle\{t\},R\rangle)$; 
the converse does not necessarily hold.

\begin{definition}\label{def:oneorkautomaton}
Let $A=\langle Q,\Sigma,\delta\rangle$
be a DTA. Let $n$ be a natural number.
We say that $A$ is a {\em $1$-or-$n$ DTA} if each state $q$
in $Q$ satisfies either $|\LL(A,q)|=1$
or $|\LL(A,q)|\geq n$.
\end{definition}

\begin{lemma}\label{lemma-transformationtosingle}
Let $\langle S,M\rangle$ be a set of terms with regular constraints.
Then, $\langle S,M\rangle$ can be transformed 
in exponential time 
into a set of terms with
single regular constraints $\langle S',R\rangle$ such that 
$\LL(\langle S',R\rangle)=\LL(\langle S,M\rangle)$ and the following
properties hold.
\begin{itemize}
\item $R=\langle A,C\rangle$ satisfies that 
$A$ is a $1$-or-$|S|$ DTA.
\item $A=\langle Q,\Sigma,\delta\rangle$ is complete and satisfies that
$|Q|\leq\|M\|^{|M|}\cdot|S|$
and $|\delta|\leq |\Sigma|\cdot |Q|^{2}=|\Sigma|\cdot \|M\|^{2|M|}\cdot|S|^{2}$
\item $|S'|\leq |S|\cdot|Q|^{|M|}\leq
  \|M\|^{|M|^2}\cdot|S|^{|M|+1}$
\item Each term in $S'$ is structurally similar to some term in $S$.
In particular, $\StructDiff(S')\leq|S|$.
\item Every two distinct terms $s,t\in S'$ are not structurally equal
with respect to $R$.
\item Each two distinct terms $s,t\in S'$ do not share variables.
\end{itemize}
\end{lemma}

\begin{proof}
Let $M=\{x_1\mapsto A_1,\ldots,x_n\mapsto A_n\}$ and 
$A_i=\langle Q_i,F_i,\Sigma,\delta_i\rangle$ for $1\leq i\leq n$.
We first complete each DTA $A_i$ to a new
DTA $A_i'=\langle Q_i',F_i,\Sigma,\delta_i'\rangle$
by adding a sink state and all undefined transitions to it.
Recall the assumption that the maximum arity of $\Sigma$ is $2$.
Thus, $|Q_i'|=|Q_i|+1$ and $|\delta_i'|=|\Sigma|\cdot
|Q_i'|^2=|\Sigma|\cdot (|Q_i|+1)^2$.
We now construct the product automaton (without accepting states) 
$A'=\langle Q',\Sigma,\delta'\rangle$, i.e., we 
set $Q'=Q_1'\times\cdots\times Q_n'$ and if,
for each $1\leq i\leq n$, 
$\delta_i'$ has the transition $f(q_{i,1},\dots,q_{i,k})\to q_i$,
then we add the transition
$f(
\langle q_{1,1},\dots,q_{n,1}\rangle,\dots,
\langle q_{1,k},\dots,q_{n,k}\rangle)
\to\langle q_1,\dots,q_n\rangle$ to
$\delta'$.
Since each state of $A'$ is a tuple of 
$|M|$ states of the automata in $M$ plus a sink state,
$|Q'|\leq\|M\|^{|M|}$.

We then transform $A'$
into a $1$-or-$|S|$ DTA. To this end, we compute 
the mapping $M':Q\to\{1,\ldots,|S|\}$ with
$M'=\operatorname{countUpto}(A',|S|)=\{(q,\Min(|\LL(A,q)|,|S|))\mid q\in Q'\}$,
according to Lemma~\ref{lem:DTA_folklore}.
Now, using $M'$ we obtain the desired $A$ as output
of the following algorithm.

\begin{small}
\begin{tabbing}
oo \= oo \= oo \= oo \= oo \= oo \kill
Input: $A'=\langle Q',\Sigma,\delta'\rangle$ and $M':Q'\to\{1,\dots,|S|\}$.\\
$Q:=\{q~\mid~ q\in Q'\wedge M'(q)=|S|\} \ \cup$ \\
\> $\{q^i ~\mid~ q\in Q' \wedge 1\leq i\leq M'(q)<|S|\}$.\\
$\delta:=\emptyset$.\\
For each $q$ in $Q'$ do:\\
\> If $M'(q)=|S|$ then:\\
\> \> For each $f(q_1,\ldots,q_m)\to q$ in $\delta'$ do:\\
\> \> \> For each $i_1,\ldots,i_m$ with $q_1^{i_1}\dots,q_m^{i_m}\in Q$ do:\\
\> \> \> \> Add $f(q_1^{i_1},\ldots,q_m^{i_m})\to q^1$ to $\delta$.\\
\> else:\\
\> \> Let $l_1\to q,\ldots,l_k\to q$ be all transitions of $\delta'$\\
\> \> \ \qquad\quad with $q$ as right-hand side.\\
\> \> counter:=1.\\
\> \> For each $i$ in $\{1,\ldots,k\}$ do:\\
\> \> \> Let $f(q_1,\ldots,q_m)\to q$ be $l_i\to q$.\\
\> \> \> For each $i_1,\ldots,i_m$ with $q_1^{i_1},\dots,q_m^{i_m}\in Q$ do:\\
\> \> \> \> Add $f(q_1^{i_1},\ldots,q_m^{i_m})\to q^{\text{counter}}$ to $\delta$.\\
\> \> \> \> counter++.\\
Complete $A=\langle Q,\Sigma,\delta\rangle$ and return the result.
\end{tabbing}
\end{small}

It is clear that this algorithm generates a complete $1$-or-$|S|$
DTA $A$ with $|Q|\leq \|M\|^{|M|}\cdot|S|$, because at most
$|S|$ new states are created for every state in $Q'$.
Moreover, since the maximum arity of $\Sigma$ is 2,
then at most $|\delta|=|\Sigma|\cdot |Q|^{2}$ transitions are
possible with such
number of states. The construction runs in exponential time because
$A'$ is constructed in exponential time, 
$M'$ is constructed in time polynomial in $|A'|$
by Lemma~\ref{lem:DTA_folklore}, and
$A$ is constructed in time $\OO(|A|)$.

Now, the set $S'$ is obtained in the following way.
Recall that the states $q$ in $Q$ are in fact of the
form $q=\langle q_1,\ldots,q_n\rangle^{j}$, i.e., are tuples
of states $q_1\in Q_1',\ldots,q_n\in Q_n'$ plus an index
$j$ satisfying $1\leq j\leq M'(\langle q_1,\ldots,q_n\rangle)$.
For each variable $x_i$ in the domain of $M$, we
define the set of variables
$V(x_i)=\{x_{\langle q_1,\ldots,q_n\rangle^{j}}^i\mid
q_i\in F_i\wedge\langle q_1,\ldots,q_n\rangle^{j}\in Q\}$.
We define the domain $V$ of the mapping $C$ as
$\bigcup_{i\in\{1,\ldots,n\}}(V(x_i))$, and the image
of each $x_q^i$ by $C$ as $q$.
Finally, let $\Theta$ be the set of substitutions $\varphi$
over $\{x_1,\ldots,x_n\}$
satisfying $\varphi(x_i)\in V(x_i)$. We compute
$S'$ as a minimal set satisfying that each one of its terms is
structurally equal to some term
in $\{\varphi(s)\mid s\in S\;\wedge\;\varphi\in\Theta\}$,
and vice-versa (i.e.\ $S'$ is computed from
$\{\varphi(s)\mid s\in S\;\wedge\;\varphi\in\Theta\}$
by removing repetitions modulo structural equality).
Moreover, we force the terms in $S'$ to do not share
variables, by renaming them in $S'$, and defining
them in $V$ and $C$ whenever it is necessary.
Obviously, each term in $S'$ is structurally
similar to some term in $S$, and any two distinct terms
in $S'$ are not structurally equal.
Each $V(x_i)$ has at most $|Q|$ variables.
Thus, $\Theta$ has at most $|Q|^{|M|}$ substitutions,
and hence $|S'|\leq |S|\cdot|Q|^{|M|}$.
Generating $S'$ consists of considering all of such
combinations of a term in $S$ and a substitution
in $\Theta$. Thus, the time complexity for creating $S'$
from $S$ and $A$ is proportional to its size, 
i.e., is in $\OO(\|S\|\cdot|Q|^{|M|})$. 
In total, $\langle S',R=\langle A,C\rangle\rangle$ is
constructed in exponential time w.r.t.
$\|S\|+\|M\|$.
\end{proof}

\subsection{Adding height constraints}\label{subsec:heightconstraints}

Let $\langle S_2,R_1\rangle$ by the set of terms with single regular
constraints that was obtained from $\langle S_1,M_1\rangle$ 
according to Lemma~\ref{lemma-transformationtosingle}.
Our algorithm proceeds by considering a term
$s$ in $S_2$, and analyzing the kind of instances which are 
in $\LL(\langle \{s\},R_1\rangle)$ but not in 
$\LL(\langle S_2-\{s\},R_1\rangle)$. 
Depending on this analysis, it either
concludes non-regularity of $\LL(\langle S_2,R_1\rangle)$, or
deduces that the height of the substitutions for some variables
of $s$ can be bounded by $|Q|+2H$, where $H$ is the maximum
height of the terms in $S_1$. 
To manage this height constraint, we extend the notion of
single regular constraint as follows.


\begin{definition}\label{def:restregconstr}
A \emph{restricted regular constraint} (\emph{over $\Sigma$})
is a tuple
$R=\langle A,V,C,W,h\rangle$, where
$W\subseteq V$ are sets of variables,
$A=\langle Q,\Sigma,\delta\rangle$ is a DTA,
$C$ is a mapping $C:V\to Q$, and
$h$ is a natural number.
The \emph{size} $\|R\|$ of $R$ is $|V|+\|A\|$.
A \emph{solution} of $R$ is
a mapping $\varphi:V\to T_\Sigma$ such
that for all $x\in V$ it holds
$\varphi(x)\in \LL(A,C(x))$,
and moreover, if $x\in W$ then $\height(\varphi(x))\leq h$.
For a finite set $S\subseteq T_\Sigma(V)$, the pair
$\langle S,R\rangle$ is 
a \emph{set of terms with restricted regular constraints}.
The language $\LL(\langle S,R\rangle)$ of $\langle S,R\rangle$ is
$\{t\mid\exists \varphi,s:(t=\varphi(s)\wedge 
s\in S \wedge\varphi\mbox{ is a solution of }R
)\}$. A term in $\LL(\langle S,R\rangle)$ is also called an
instance of $\langle S,R\rangle$.
\end{definition}

Obviously, the set of terms with single regular constraints 
$\langle S_2,R_1=\langle A,C\rangle\rangle$ 
can be transformed into the set of
terms with restricted regular constraints
$\langle S_2,R_2=\langle A,V,C,\emptyset,|Q|+2H\rangle\rangle$,
and the represented language is preserved, i.e.
$\LL(\langle S_2,R_1\rangle)=\LL(\langle S_2,R_2\rangle)$.
For a restricted regular constraint $\langle S,R\rangle$,
we can define the infinite-instances property analogously to
Definition~\ref{def:infinst}, where it is defined for a set of
terms with regular constraints. As mentioned before, when a
term in $S$ satisfies the infinite-instances property, then 
$\LL(\langle S,M\rangle)$ is not regular~\cite{godmantis08}. Exactly
the same thing, with the same proof, can be said about a set of
terms with restricted regular constraints $\langle S,R\rangle$.

\begin{lemma}\label{lem:nonreg}
Let $\langle S,R\rangle$ be a set of terms with
restricted regular constraints.
Let $s$ be a term satisfying the infinite-instances property
in $\langle S,R\rangle$.
Then, $\LL(\langle S,R\rangle)$ is not regular.
\end{lemma}

In order to make the paper self-contained, we prove this result.
The proof is simplified and adapted to the
case of restricted regular constraints.

\begin{proof}
We prove the lemma by contradiction, i.e.\ we assume
that there exists DTA
$B=\langle Q_B,\Gamma,\delta_B,F_B\rangle$ recognizing
$\LL(\langle S,R\rangle)$ in order to reach a contradiction.

By the assumptions, there exists a variable $x$
with more than one occurrence in $s$,
and infinite instances $\varphi_1(s),\varphi_2(s),\ldots$ of 
$\langle \{s\},R\rangle$ which are not instances of
$\langle S-\{s\},R\rangle$,
and satisfying $\varphi_i(x)\not=\varphi_j(x)$ for all $j>i\geq 1$.

Let $R$ be $\langle A,V,C,W,h\rangle$, let
$A$ be $\langle Q_A,\Gamma,\delta_A\rangle$,
and let $H$ be the maximum height of the terms in $S$.
Let $p_1$ be one of the positions in $s$ where $x$ occurs.

Since the instances $\varphi_i(s)$ are not in
$\langle S-\{s\},R\rangle$ and are different on $x$,
there is a solution $\varphi$ ($\varphi=\varphi_i$ for some $i\geq 1$)
of $R$ satisfying that $\varphi(s)$
is not an instance of $\langle S-\{s\},R\rangle$
and $\height(\varphi(x))>H+h+|Q_A|\cdot|Q_B|$.
Let $p_2$ be a position such that $p_1.p_2$ is
a position of $\varphi(s)$, $|p_1.p_2|=H+h$ and
$\height(\varphi(s)/(p_1.p_2))>|Q_A|\cdot|Q_B|$.
By a simple pumping argument, there exist positions
$p_3$ and $p_4$ satisfying that $p_1.p_2.p_3.p_4$
is a position of $\varphi(s)$, $|p_4|\geq 1$,
$A(\varphi(s)/(p_1.p_2.p_3.p_4))=
A(\varphi(s)/(p_1.p_2.p_3))$
and
$B(\varphi(s)/(p_1.p_2.p_3.p_4))=
B(\varphi(s)/(p_1.p_2.p_3))$.

Let $\overline{H}$ be $\height(\varphi(s))$.
Let $D$ be the context $(\varphi(s)/(p_1.p_2.p_3))[p_4\from \bullet]$.
We consider the term
$t=\varphi(s)[p_3\from D^{\overline{H}}[\varphi(s)/p_4]]$.
Note that $t$ is accepted by $B$. Thus, in order to
reach a contradiction, it suffices to see that
$t$ is not an instance of $\langle S,R\rangle$.
It is clearly not an instance of $\langle s,R\rangle$,
since we have the term
$\varphi(s)[p_3\from D^{\overline{H}}[\varphi(s)/p_4]]/p_1$
as a subterm in $t$
at a position of $x$ in $s$, and the term
$\varphi(s)/p_1$ as a subterm in $t$ at another position of $x$ in $s$.
Thus, it rests to see that $t$ is not an instance
of $\langle \{s'\},R\rangle$ for each $s'$ in $S-\{s\}$.

For each term $s'$ in $S-\{s\}$, we know that the term
$\varphi(s)$ is not an instance of $\langle \{s'\},R\rangle$,
and this has to be due to one of the following reasons:
\begin{itemize}
\item[(a)] There is a position
$q$ in $\Posnv(s')$ satisfying that
$q$ is not in $\Pos(\varphi(s))$,
\item[(b)] There is a position
$q$ in $\Pos(\varphi(s))\cap \Posnv(s')$
satisfying $\varphi(s)[q]\not=s'[q]$,
\item[(c)] There is a position $q$
in $\Pos(\varphi(s))\cap\Posv(s')$ satisfying
$A(\varphi(s)/q)\not=C(s'[q])$,
\item[(d)] There are positions $q$ and $q'$
in $\Pos(\varphi(s))\cap\Posv(s')$
satisfying $s'[q]=s'[q']$ and
$\varphi(s)|_{q}\not=\varphi(s)|_{q'}$.
\item[(e)] There is a position $q$
in $\Pos(\varphi(s))\cap\Posv(s')$
satisfying $s'[q]\in W$ and $\height(\varphi(s))>h$.
\end{itemize}

In cases (a), (b), (c) and (e) it is straightforward that
$t$ is not an instance of $\langle \{s'\},R\rangle$ by the same reason.
Thus, assume we are in case (d). If both $q$ and $q'$ are disjoint
with $p_1.p_2$, then $t/q=\varphi(s)/q\not=\varphi(s)/q'=t/q'$, and hence,
$t$ is not an instance of $\langle \{s'\},R\rangle$. If one
of $q$ or $q'$, say $q$, is a prefix of $p_1.p_2$,
then, $t/q\not=t/q'$ also holds, because
$\height(t/q)>\overline{H}\geq\height(t/q')$.
Therefore, $t$ is not
an instance of $\langle \{s'\},R\rangle$ in any case,
and this concludes the proof.
\end{proof}

For the particular case of a singleton $S=\{s\}$,
Lemma~\ref{lem:nonreg} implies the following statement.

\begin{corollary}
Let $\langle\{s\},R\rangle$ be a set of terms with 
restricted regular constraints.  
Then, $\LL(\langle \{s\},R \rangle)$ is
regular if and only if for each variable $x$ occurring at least twice
in $s$, either $|\LL(A,C(x))|\not=\infty$ or $x\in W$.
\end{corollary}

The previous corollary naturally leads to the following definition
of regular term.

\begin{definition}
Let $R=\langle A,V,C,W,h\rangle$ be a restricted regular
constraint.
A term $s\in T_\Sigma(V)$
is {\em regular with respect to $R$} if for each variable
$x$ occurring at least twice in $s$,
either $|\LL(A,C(x))|\not=\infty$ or $x\in W$.
\end{definition}

\subsection{Determining a term}\label{sec:determining-term}

At this point, we want to test whether a term $s$ satisfies the
infinite-instances property with respect to $S_2$, that is,
we want to analyze the instances
of $\langle \{s\},R_2\rangle$ which are not instances 
of $\langle S_2-\{s\},R_2\rangle$. 
To make this problem easier, it would
be good to have $s$ determined at all non-variable positions of the
terms in $S_2$, according to the following definition.

\begin{definition}\label{def:determined}
For a position $p$ and a term $s\in T_\Sigma(V)$, 
we say that $s$ is {\em determined} at $p$ if
either $p\in\Posnv(s)$ 
or there is a prefix $p'$ of $p$ such 
that $s[p']$ is a constant symbol, i.e., it is in $\Sigma^{(0)}$.
The term $s$ is determined at a set of positions $P$
if it is determined at each $p\in P$.
\end{definition}

One of the nice (and obvious) properties of determined positions $p$
of $s$ is that, for any substitution $\varphi$ mapping variables to terms,
the symbol $\varphi(s)[p]$ is either undefined or coincides with $s[p]$.

\begin{lemma}
Let $p$ be a position and $s$ a term determined at $p$.
Let $\varphi_1,\varphi_2$ be mappings from variables to $T_\Sigma$. 
Either $p$ is not a position of both $\varphi_1(s)$ and
$\varphi_2(s)$, or $\varphi_1(s)[p]=\varphi_2(s)[p]$.
\end{lemma}

\begin{proof}
No prefix $p'$ of $p$ is such that $s[p']$ is a variable. Hence, for
every substitution $\varphi$, we have that $\varphi(s)$ is undefined
at $p$ if so was $s$, or that $\varphi(s)[p]=s[p]$ if not.
\end{proof}

Another nice property of determined positions is that,
given a term $s$, a restricted regular constraint $R$,
and a set of positions $P$, then, a set of terms
$s_1,\ldots,s_k$, all of them determined at $P$, can be generated in 
exponential time
on $|\Prefixes(P)|$,
such that $\{s\}$ and $\{s_1,\ldots,s_k\}$
represent the same language. The idea of determining
a term at a set of positions was already used in~\cite{godmantis08}.

\begin{lemma}\label{lemma-determining}
Let $R=\langle A,V,C,W,h\rangle$ be a restricted regular constraint,
where $A=\langle Q,\Sigma,\delta\rangle$.
Let $s$ be a term in $T_\Sigma(V-W)$ and let $P$ be a set of positions.
It  can be computed in time
$\OO(|s|\cdot |\Prefixes(P)|\cdot \|R\|^{|\Prefixes(P)|})$
an extension $R'=\langle A,V',C',W,h\rangle$ of $R$ and
a set of terms $\{s_1,\ldots,s_k\}$ in $T_\Sigma(V')$
satisfying the following properties.
\begin{itemize}
\item $s_1,\dots,s_k$ are determined at $P$.
\item
$\LL(\langle\{s_1,\ldots,s_k\},R'\rangle)=\LL(\langle\{s\},R\rangle)$.
\item 
Each $s_i$ can be obtained from $s$ 
through a substitution which replaces each variable by a
term with height bounded by the maximum length 
of a position in $P$.
\item $k\leq|\delta|^{|\Prefixes(P)|}$.
\item For each $i$ in $\{1,\ldots,k\}$, 
$|s_i|$ is bounded by
$3\cdot |\Prefixes(P)|\cdot |s|$.
\end{itemize}
\end{lemma}

\begin{proof}
We start with $\langle S_1',R_1'\rangle=\langle \{s\},R\rangle$ and
transform it iteratively, while preserving the represented language,
into new pairs
$\langle S_2',R_2'\rangle,\dots,\langle S_f',R_f'\rangle$, where
we denote $S_f'=\{s_1,\dots,s_k\}$ and $R_f'=R'$. 
Let $s'$ be a term of $S_i'$ which is not determined at some
$p\in \Prefixes(P)\cap\Posv(s')$, and let
$R_i' = \langle A,V_i,C_i,W,h \rangle$ be the $i$-th constraint.
Let $y=s'[p]$ and let 
$q=C_i(y)$.
The DTA $A$ has a
finite number of transitions of the form
$g(q_1,\ldots,q_m)\rightarrow q$, where 
$q,q_1,\ldots,q_m\in Q$ and $g\in\Sigma^{(m)}$ for $m\leq 2$, by the
assumption on $\Sigma$.
For each such transition, we construct the 
substitution $\gamma_{g,q_1,\ldots,q_m,q}=[y\leftarrow g(z_1,\ldots,z_m)]$
where $z_1,\ldots,z_m$ are new variables. Let $V_{i+1}'$ be
the union of these new sets of variables for all
such transitions. Let $C_{i+1}'$ be the union
of all sets $\{(z_1,q_1),\dots,(z_m,q_m)\}$ for all
such transitions.
We set $V_{i+1} := V_{i+1}'\cup V_i$ and
$C_{i+1}:=C_{i+1}'\cup C_i$.
Finally, we set
$S_{i+1}' := (S_{i}'-\{s'\})\cup S''$, where $S''$ is the set of terms
obtained by applying all the substitutions
$\gamma_{g,q_1,\ldots,q_m,q}$ to $s'$. 
Clearly, $\LL(\langle S_{i+1}',R_{i+1}'\rangle)$ coincides with 
$\LL(\langle S_i,R_i\rangle)$.

At each of the $|\Prefixes(P)|$ positions we apply at most $|\delta|$
different substitutions giving us at most $|\delta|^{|\Prefixes(P)|}$-many
different terms. Thus, $k\leq|\delta|^{|\Prefixes(P)|}$.
Each substitution $\gamma_{g,q_1,\ldots,q_m,q}$ increases the size of
a term $s'$ by the arity $m$ of $g$, which is at most $2$, and the
variable replaced has at
most $|s|$ occurrences
in $s'$. Thus, $|s|+|\Prefixes(P)|+2\cdot |\Prefixes(P)|\cdot |s|\leq
3\cdot |\Prefixes(P)|\cdot |s|$
bounds the size of each $s_i$.

Note that only those variables $y$ which appear at some
position $p\in P$ may be replaced by some
$\gamma_{g,q_1,\ldots,q_m,q}=[y\leftarrow g(z_1,\ldots,z_m)]$.
Since the new variables $z_i$ always appear one position deeper than
the variable $y$ they substitute, it follows that, in the
process described, no variable of $s$ can be replaced by
a term of height larger than the maximum length of
the positions in $P$.
\end{proof}

\subsection{Structurally subsumed terms}\label{sec:subsumed-terms}

Let $s$ be a term determined at all the non-variable positions of the terms
in $S_2$. In order to check the infinite-instances property,
 our goal is to characterize the set $\LL(\langle
 \{s\},R_2\rangle)- \LL(\langle S_2-\{s\},R_2\rangle)$. Recall that
 $\StructDiff(S_2)$ is bounded by the initial $|S_1|$. This can be 
used to discard many terms in $S_2$ having no common instance with
$\langle \{s\},R_2\rangle$. To this end, we introduce the following notions.
Let $A$ be a DTA and $C$ a mapping from
variables to states of $A$. 
For a term $s\in T_\Sigma(V)$, we define 
$C(s):=A(s[x\leftarrow C(x)\mid x\in V])$.
If $C$ is clear from the context, we denote
a term $s$ by $s^q$ for $q=C(s)$,
or as $f^q(s_1,\ldots,s_m)$ if $s=f(s_1,\ldots,s_m)$.

\begin{definition}\label{def:structsub}
Let $R=\langle A,V,C,W,h\rangle$ be a
restricted regular constraint over $\Sigma$,
and let $s,t\in T_\Sigma(V)$.
We say that $s$ is {\em structurally subsumed} by $t$ (with respect to
$R$), 
if for all $p$ in $\Posnv(t)$
it holds that $p$ is in $\Pos(s)$ and $t[p]=s[p]$, and moreover,
for all $p$ in $\Pos(t)$ it holds that $C(t/p)=C(s/p)$.
\end{definition}

Which terms in $\LL(\langle S_2-\{s\},R_2\rangle)$ can possibly have
common instances with $s$? 
If $t$ structurally subsumes $s$, then they potentially have
common instances (this depends on the equality constraints imposed
by duplicated variables in $s$ and $t$). For instance, 
$t=f(x,x)$ structurally subsumes $s=f(a,b)$ if 
$C(x)=C(a)=C(b)$, but obviously $s$ and $t$ do not have common
instances. What happens if $t$ does \emph{not} structurally subsume
$s$? Does this imply that $s$ and $t$ do not have common instances?
Unfortunately not: $t=f(x,a)$ does \emph{not} structurally subsume
$s=f(a,y)$, but if $C(y)=C(x)$ and $a\in\LL(A,C(x))$
then $\langle \{t\},R\rangle$ and $\langle\{s\},R\rangle$
share $f(a,a)$ as instance.
At this point, the benefits of determining a
term come into play.

\begin{lemma}\label{lemma-nonsubsumesimpliesdisjoint}
Let $R=\langle A,V,C,W,h\rangle$ be a restricted regular constraint
over $\Sigma$ and let $s,t\in T_\Sigma(V)$.
If $s$ is determined at $\Posnv(t)$ and
$s$ is not structurally subsumed by $t$, then
$\LL(\langle \{s\},R\rangle)$ and 
$\LL(\langle \{t\},R\rangle)$ are disjoint.
\end{lemma}

\begin{proof}
With the conditions of the lemma, and according to
Definition~\ref{def:structsub}, either it exists a position $p\in\Posnv(t)
\subseteq \Posnv(s)$ such that $t[p]\neq s[p]$,
or it exists a position $p\in \Pos(t)\subseteq \Pos(s)$ such
that $C(t/p)\neq C(s/p)$. In the former case it is clear that all
instances of $\langle \{s\},R\rangle$ and $\langle
\{t\},R\rangle$ differ at $p$; in the latter case, the result follows
from the fact that $\LL(A,q)$ and $\LL(A,q')$ are disjoint if
$q\neq q'$.
\end{proof}

Moreover, when two terms are structurally similar but not structurally
equal, they cannot both structurally subsume a third term.

\begin{lemma}\label{lemma-limitedstructurallysubsumes}
Let $R=\langle A,V,C,W,h\rangle$ be a restricted regular constraint
over $\Sigma$ and $s, t_1, t_2\in T_\Sigma(V)$.
Assume that $t_1$ and $t_2$
are structurally similar but not structurally equal, and that
$s$ is structurally subsumed by $t_1$.
Then $s$ is not structurally subsumed by $t_2$.
\end{lemma}
\begin{proof}
If two terms $t_1$ and $t_2$ are structurally similar but not
structurally equal, then there is a position 
$p\in\Posv(t_1)=\Posv(t_2)$ 
such that $C(t_1[p])\neq C(t_2[p])$.
Now $t_1$ structurally subsumes $s$, so $C(t_1[p]) = C(s[p])$;
this prevents $t_2$ from subsuming $s$.
\end{proof}

Recall that, by Lemma~\ref{lemma-transformationtosingle}, we can
choose at most $|S_1|$ non-structurally similar terms in $S_2$.  This
fact, combined with Lemma~\ref{lemma-limitedstructurallysubsumes},
implies that at most $|S_1|-1$ terms in $S_2-\{s\}$ structurally
subsume $s$. Since, by assumption, $s$ is determined at all
non-variable positions of terms in $S_2$, then, by
Lemma~\ref{lemma-nonsubsumesimpliesdisjoint}, only those $|S_1|-1$
terms may have common instances with $s$.  Thus, when analyzing the
instances of $\langle \{s\},R_2\rangle$ which are not instances of
$\langle S_2-\{s\},R_2\rangle$, we can first choose the subset $S_3$
of terms in $S_2-\{s\}$ which structurally subsume $s$ (because they
are the only possible ones to have common instances with $s$), and study
which instances of $\langle \{s\},R_2\rangle$ are not instances of
$\langle S_3,R_2\rangle$. Note that $|S_3|\leq |S_1|-1$.

As mentioned before, if $t$ structurally subsumes $s$, then 
whether they have common instances or not, depends on the
equality constraints imposed by duplicated variables. Since our
restricted regular constraints also require that 
$\varphi(x)\leq h$ for $x\in W$, it means that $\varphi(s)$ can only
be an instance of $t$ if the height of $\varphi(s)/p$ is 
smaller than or equal to $h$ whenever $t[p]\in W$.

\begin{lemma}\label{lemma-structurally-subsumed}
Let $R=\langle A,V,C,W,h\rangle$ be a restricted regular constraint.
Let $s$ and $t$ be terms such that $s$ is structurally subsumed
by $t$ with respect to $R$, and $\Vars(s)\cap W=\emptyset$.
Let $\varphi(s)$ be an instance of $\langle\{s\},R\rangle$.
Then $\varphi(s)$ is an instance of $\langle\{t\},R\rangle$ if
and only if
\begin{itemize}
\item for all $p,q$ in $\Posv(t)$ 
such that $t[p]=t[q]$ it holds
$\varphi(s)/p=\varphi(s)/q$, and 
\item for all $p$ in $\Posv(t)$ such that $t[p]\in W$
it holds $\height(\varphi(s)/p)\leq h$.
\end{itemize}
\end{lemma}
\begin{proof}
Since $s$ is structurally subsumed by $t$,
an instance $\varphi'(t)$ of $t$ coincides with $\varphi(s)$
if and only if $\varphi'(t)/p=\varphi(s)/p$
for every $p$ in $\Posv(t)$. But this condition
uniquely determines $\varphi'$, i.e.\ $\varphi(s)$ is an instance
$\varphi'(t)$ of $t$ if and only if $\varphi'$
is defined as $\varphi'(t[p]):=\varphi(s)/p$, for every $p$ in 
$\Posv(t)$. This definition of $\varphi'$ is correct
(i.e.\ uniquely defined for each variable) if and only if for all
$p,q$ in $\Posv(t)$ such that $t[p]=t[q]$ it holds
$\varphi(s)/p=\varphi(s)/q$. Thus, the first item
is necessarily satisfied.
Moreover, by the assumptions of the lemma,
for all $p\in\Pos(t)$ it holds $C(t/p)=C(s/p)$.
Thus, the instance $\varphi'(t)$ of $t$ is also an
instance of $\langle t,R\rangle$ if and only if
for all $p\in\Posv(t)$ such that $t[p] \in W$, it holds
$\height(\varphi(s)/p)\leq h$, as required by the second item of
the lemma.
\end{proof}

\subsection{Formulas representing instances}\label{sec:formula-instances}

By using Lemma~\ref{lemma-structurally-subsumed}, we are able to characterize
the instances of $\langle \{s\},R_2\rangle$ which are not instances
of $\langle S_3,R_2\rangle$ as
the solutions of a formula $F$ which is
a disjunction of conjunctions with
inequalities between terms and height restrictions of terms
as predicates, and a single regular constraint for the variables.

\begin{definition}
Let $V$ be a finite set of variables.
A {\em formula with inequality and height predicates} $F$
(over $V$)
is a disjunction of conjunctions of predicates of the form
$s\not= t$ and $\height(s)> h$, where $s,t\in T_\Sigma(V)$
and $h$ is a natural number.
A {\em constrained formula} of order $n$ is a triple
$\langle F,A,C\rangle$, where $A=\langle Q,\Sigma,\delta\rangle$ is a
$1$-or-$(n+1)$ DTA, 
$F$ is a formula with inequality and
height predicates where every conjunction has at most $n$ predicates,
and $C$ is a total function $C:V\to Q$. Moreover, 
for each predicate $\height(s)> h$ we require 
that $h$ is greater than or equal to $|Q|+\height(s)$.
A {\em solution} of $\langle F,A,C\rangle$ is a
substitution $\varphi:V\to T_\Sigma$ such that 
$A(\varphi(x))=C(x)$ for each $x\in V$ and
$\varphi(F)$ evaluates to true by interpreting $\not=$, $\height$,
and $>$ in the natural way. The set of all solutions is denoted
$\Sol(\langle F,A,C\rangle)$. 
\end{definition}

We now construct a constrained formula for a given 
set of terms $S$ and term $s$.
Denote by $\selPos(S)$ the set of functions
$P:S\to\cup_{s'\in S}(\Posv(s'))$ such that
for each $s'\in S$, $P(s')\in\Posv(s')$.
Let $W$ be a set of variables. We define $\selPos(S,W)$ as the
subset of functions $P$ of $\selPos(S)$
such that for each $s'\in S$, $s'[P(s')]\in W$.

\begin{definition}\label{definition-formula}
Let $S$ be a set of terms, and let $R=\langle A,V,C,W,h\rangle$
be a restricted regular constraint,
where $A=\langle Q,\Sigma,\delta\rangle $ is a $1$-or-$(|S|+1)$ DTA.
Finally, let $s\in T_\Sigma(V-W)$ be a term that is structurally
subsumed by all terms in $S$ with respect to $R$.
Also suppose that $h$ is greater than or equal to $|Q|+\height(s)$.
We define ${\mathcal F}(s,S,W,h)$ as
$$
\bigvee_{\alpha}
\left(\bigwedge_{t\in S'}s/P(t)\not=s/U(t)
\bigwedge_{t\in S-S'}\height(s/T(t))>h
\right)
$$
where $\alpha$ says that 
$S'\subseteq S$; $P,U\in\selPos(S')$
such that
for every $s'\in S'$: 
$P(s')\not=U(s')$ and $s'[P(s')]=s'[U(s')]$;
and
$T\in\selPos(S-S',W)$.
Note that ${\mathcal F}(s,S,W,h)$ is a formula with inequality
and height predicates and that
$\langle {\mathcal F}(s,S,W,h),A,C\rangle$ is a constrained formula
of order $|S|$.
\end{definition}


According to Lemma~\ref{lemma-structurally-subsumed}, 
the instances of $\langle \{s\},R\rangle$ that are \emph{not} instances of
$\langle S,R\rangle$ are precisely the terms $\varphi(s)$ with
$\varphi\in\Sol(\langle {\mathcal F}(s,S,W,h), A, C\rangle)$.
We state this in the following lemma.

\begin{lemma}
Let $R=\langle A,V,C,W,h\rangle$ be a restricted regular constraint,
where $A$ is $\langle Q,\Sigma,\delta\rangle$.
Let $S$ be a set of terms and let $s\in T_\Sigma(V-W)$ be a term that is structurally
subsumed by all terms in $S$,
and such that $h\geq |Q|+\height(s)$. Then,
$\LL(\langle \{s\},R\rangle)-\LL(\langle S,R\rangle)=
\Sol(\langle {\mathcal F}(s,S,W,h), A, C\rangle)$.
\end{lemma}

Our goal is to decide whether $\langle \{s\}, R_2\rangle$ has infinitely
many instances, all of them different on a certain variable $x$, 
and all of them 
not instances of $\langle S_3, R_2\rangle$. We do not solve this
problem for an arbitrary restricted regular constraint $R$.
Recall that $|S_3|\leq |S_1|-1$, $R_2$ is of the form
$\langle A,V,C,W,|Q|+2H)\rangle$ where $H$ is the maximum height of
a term in $S_1$, $\height(s)\leq 2H$
and $A$ is a $1$-or-$|S_1|$ DTA.
Our problem now translates to the constrained formula
$\langle {\mathcal F}(s,S_3,W,|Q|+2H),A,C\rangle$, which is of order
$|S_3|$ due to the particularities of $R_2$; i.e.,
we need to decide
whether $\langle {\mathcal F}(s,S_3,W,h),A,C\rangle$
has infinite solutions and all of them
different on a concrete variable $x$.
To this end we proceed by
transforming this formula by means of the set of rules described in
Figure~\ref{table-one}. The following lemma states that
the inference system preserves the set of solutions.

\newcommand{\ttrule}[4]{{\bf {\hspace{-0.2cm}#1}}: & $\begin{array}{c}{#2}\\ \hline {#3}\end{array}$ \\ 
 \multicolumn{2}{p{3.2in}}{{#4}}}

\begin{figure*}[ht]
\begin{center}

\begin{tabular}{lc}

\ttrule{{Remove-insat1}}%
{C \vee (t\not=t \wedge D)}%
{C}%
{}

\\*[0.5em]

\ttrule{{Remove-insat2}}%
{C \vee (s^q\not=t^q \wedge D)}%
{C}%
{where $|\LL(A,q)|=1$.}

\\*[1em]

\ttrule{{Remove-sat1}}%
{C \vee (s^q\not=t^{q'}\wedge D)}%
{C \vee (D)}
{where either $q\not=q'$, or $s[\varepsilon]$, $t[\varepsilon]$ are
not variables and $s[\varepsilon]\not=t[\varepsilon]$.}

\\*[2em]

\ttrule{{Remove-sat2}}%
{C \vee (x^q\not=t^{q}\wedge D)}%
{C \vee (D)}
{where $t$ is not $x$ and $x\in\Vars(t)$.}

%

\\*[1.5em]

\ttrule{{Decompose}}%
{C \vee (f^q(s_1,\dots,s_m)\!\! \not=\!\! f^q(t_1,\dots,t_m)\wedge D)}%
{C \vee \bigvee_{i\in\{1,\ldots m\}}(s_i\not=t_i\wedge D)}
{}

\\*[2em]

\ttrule{{Decrease-height}}%
{C \vee (\height(f^q(s_1,\ldots,s_m))>h\;\wedge D)}%
{C \vee \bigvee_{i\in\{1,\ldots m\}}(\height(s_i)>h-1 \wedge D)}
{\vspace{0.5ex} where $\LL(A,q)$ is infinite, and $h>|Q|$.}

\\*[3.5em]

\ttrule{{Remove-height}}%
{C \vee (\height(s^q)>h\;\wedge D)}%
{C}
{where $\LL(A,q)$ is finite, and $h\geq |Q|$.}

%
\end{tabular}

\end{center}
\caption{Inference rules for transforming formulas into final formulas.}
\label{table-one}
\end{figure*}

\begin{lemma}\label{lem:inference}
If $\langle F,A,C\rangle$ is a constrained formula of order $n$ and
$\langle F,A,C\rangle$ derives into $\langle G,A,C\rangle$ by the application
of one inference rule of Figure~\ref{table-one}, then 
$\langle G,A,C\rangle$ is a constrained formula of order $n$, and
$\Sol(\langle G,A,C\rangle)=\Sol(\langle F,A,C\rangle)$.
\end{lemma}

\begin{proof}
It is clear that $\langle G,A,C\rangle$ is also a constrained formula
of the same order than $\langle F,A,C\rangle$: 
the DTA does not change;  the number of predicates
in a conjunction is never increased; and the only rule that adds a
new height predicate, i.e., rule \emph{Decrease-height}, reduces both
by one the height of the left side term and the right side bound.

To see that the solutions are preserved under the applications of the
inference rules, we only need to observe that rules
\emph{Remove-insat1}, \emph{Remove-insat2}, and \emph{Remove-height}
simply remove conjunctions that are impossible to satisfy; that rules
\emph{Remove-sat1} and \emph{Remove-sat2} remove inequality statements
from inside a conjunction which are always satisfied; and that rules
\emph{Decompose} and \emph{Decrease-height} decompose a statement into
an equivalent disjunction of statements.
\end{proof}

\begin{definition}
A constrained formula $\langle F,A,C\rangle$
is {\em final} if no rule can be applied on $\langle F,A,C\rangle$.
\end{definition}

The following lemma characterizes final formulas.

\begin{lemma}\label{lem:finalformulas}
Let $\langle F,A,C\rangle$ be a constrained formula of order $n$.
Then, $\langle F,A,C\rangle$ is a final formula of order $n$
if and only if $F$ is a
disjunction of conjunctions of the form
\[(x_1\not=t_1\wedge \ldots\wedge x_m\not=t_m
\wedge
\height(y_1)>h_1\wedge\ldots\wedge\height(y_k)>h_k)
\]
where $k+m\leq n$, every $x_i$ is a variable not occurring in the
corresponding $t_i$, every $C(x_i)$ coincides with its
corresponding $C(t_i)$, every $|\LL(A,C(x_i))|>n$, and
every $y_i$ is a variable satisfying $|\LL(A,C(y_i))|=\infty$.
\end{lemma}
\begin{proof}
The right-to-left implication trivially follows by inspecting
that no inference rule can be applied on $F$.
For the left-to-right implication, assume that $\langle F,A,C\rangle$
is a final formula.
First, let $s\not=t$ be any inequality predicate of $F$.
The terms $s$ and $t$ are different since rule Remove-insat1
is not applicable. One of both has to be a variable: otherwise,
one of Remove-sat1 or Decompose is applicable. Without loss
of generality, let $s$ be a variable $x$. Then, $x$ cannot
occur in $t$: otherwise, rule Remove-sat2 is applicable
(recall that $x=s$ and $t$ are different). The states
$C(x)$ and $C(t)$ coincide: otherwise, rule Remove-sat1
is applicable. Moreover, $|\LL(A,C(x))|>n$:
since $A$ is a $1$-or-$(n+1)$ DTA,
$|\LL(A,C(s))|$ is either $1$ or greater than $n$,
but it cannot be $1$ because, otherwise, rule Remove-insat2
would be applicable.
Second, let $\height(u)>|Q|+h$ be any height predicate of $F$.
The cardinality of $\LL(A,C(u))$ must be infinite: otherwise
rule Remove-height is applicable. Moreover, the term $u$
must be a variable: otherwise, rule Decrease-height is applicable.
\end{proof}

The following lemma proves that any non-empty final formula
of order $n$
has a solution, and moreover, if some variable $x$ has an infinite
language, then there are infinitely many solutions all of them different on $x$.

\begin{lemma}\label{lemma:infsol} 
Let $V$ be a set of variables.
Let $\langle F,A,C\rangle$ be any non-empty final
formula (over $V$) of order $n$. Then, $\langle
F,A,C\rangle$ has a solution. Moreover, if $x\in V$ satisfies that
$|\{\LL(A,C(x))\}|=\infty$, then there exists infinitely many solutions
$\varphi_1,\varphi_2,\ldots$ of $\langle F, A, C\rangle$ such that all
$\varphi_1(x),\varphi_2(x),\ldots$ are pairwise different.
\end{lemma}

\begin{proof}
Note that $F$ is a disjunction of conjunctions
$\bigvee G_i$, where each $\langle G_i,A,C\rangle$
is also a non-empty final formula, and
$\Sol(\langle G_i,A,C\rangle)\subseteq\Sol(\langle
F,A,C\rangle)$ holds.
Thus, we assume the simple case where $F=G_1$ is a single conjunction of
predicates
$(x_1\not=t_1\wedge \ldots\wedge x_m\not=t_m \wedge
\height(y_1)>h_1\wedge\ldots\wedge\height(y_k)>h_k)$.

We construct a solution $\varphi$ of $\langle F,A,C\rangle$ by
first defining $\varphi(x)=t$ for each variable
$x$ satisfying that $\LL(A,C(x))$ is a singleton language $\{t\}$
(note that this is the only possible election for $\varphi(x)$
in a solution). Then, we replace all occurrences of $x$ by
$t$ in $F$. By Lemma~\ref{lem:finalformulas}, 
each occurrence of $x$ must be at a child position
of some node in a $t_i$.
After that, the resulting $F$ satisfies that
no variable $x$ with $|\LL(A,C(x))|=1$ occurs in $F$,
but the left-hand sides of inequalities are still variables,
and each $x_i\not=t_i$ satisfies that $x_i$ does not occur
in $t_i$.

Now, we complete the definition of $\varphi$ by
applying the process explained below.
This process chooses a particular variable $x$ at each step, chooses
a particular substitution $\varphi(x)$ for it, and then replaces all
occurrences of $x$ in $F$
by $\varphi(x)$. Thus, the process terminates. Since $F$ is
modified along the execution, it
can lose the property of being a final formula:
for example, when an equation $x=t$ occurs and $x$
is instantiated, it is no longer true that each equation
has a variable in one of its sides. The election
of each $\varphi(x)$ for each corresponding $x$ is
done in a way such that, whenever a predicate is
made variable-free, then it is trivially true.


\begin{itemize}
\item[(a)] If all inequalities of $F$ with variables contain at least
  two distinct variables, then choose any variable $x$ occurring in
  them. Choose any variable-free term $t$ in $\LL(A,C(x))$, also satisfying
  $\height(t)>h$ if a predicate $\height(x)>h$
  occurs in $F$ (note that such a $t$ exists since,
  by Lemma~\ref{lem:finalformulas}, the language $\LL(A,C(x))$ is infinite).
  Then, define
  $\varphi(x):=t$. Replace each occurrence of $x$ by $t$ in $F$. Jump to (a).
\item[(b)] If $F$ still contains a variable in some inequality, then
  choose an inequality $s_i\not=t_i$ with occurrences of just one
  variable $x$, i.e.\ satisfying $\Vars(s_i)\cup\Vars(t_i)=\{x\}$. Without loss of generality, let
  $s_1\not=t_1,\cdots,s_{m'}\not=t_{m'}$ be the inequalities containing $x$ and
  no other variable than $x$.
  Choose a variable-free term $t$ in $\LL(A,C(x))$, such that $\{x\mapsto
  t\}(s_1\not=t_1\wedge\cdots\wedge s_{m'}\not=t_{m'})$ is true (note that
  this is possible since $|\LL(A,C(x))|>n\geq m\geq m'$,
  where $n$ is the order
  of the final formula $\langle F,A,C\rangle$), and also satisfying
  $\height(t)>h$ if a predicate $\height(x)>h$ occurs
  in $F$ (as before, such a $t$ exists since in this case $\LL(A,C(x))$
  is infinite).  Replace each occurrence of $x$ by $t$. 
  Jump to (a).
\item[(c)] For each variable $x$ for which $\varphi$ is still not
  defined, choose any term $t$ in $\LL(A,C(x))$ also satisfying
  $\height(t)>h$ if a predicate $\height(x)>h$
  occurs in $F$, and define $\varphi(x):=t$.  Replace each
  occurrence of $x$ by $t$.
\end{itemize}

For the case of variables $x$ with infinite $\LL(A,C(x))$,
when the process above chooses a value for them, it has
an infinite number of possibilities. Hence, infinitely
many solutions $\varphi$ can be found, all of them distinct on
$\varphi(x)$.
\end{proof}

\begin{lemma} \label{lem:infiniteinstancesalgorithm}
Let $s\in T_\Sigma(V-W)$ be a term determined at $\Posnv(S)$,
where $\langle S,R\rangle$ is a set of terms with restricted regular
constraints. 
Let $R$ be of the form
$\langle A,V,C,W,h\rangle$, where $A$ is a $1$-or-$(|S|+1)$ DTA
and $h\geq |Q|+\height(s)$.
It is decidable in time $\OO(2^{|S|}\cdot |s|^{2|S|+1}|S|)$ whether
$\langle\{s\},R\rangle$ has an instance not in $\LL(\langle S,R\rangle)$.
In the affirmative case, if a variable $x$
occurs at least twice in $s$ and it satisfies
$|\LL(A,C(x))|=\infty$, then $s$ has infinitely many instances
not in $\LL(\langle S,R\rangle)$, and all of them
different on $x$.
\end{lemma}

\begin{proof}
Let $F$ be $\mathcal{F}(s,S,W,h)$. The constrained formula
$\langle F,A,C\rangle$ of order
$|S|$ can be easily constructed in time $T=2^{|S|+1}\cdot |S|\cdot
|s|^{2|S|+1}$, since it has no more than $2^{|S|}\cdot |s|^{2|S|}$
conjunctions, each of them with at most $|S|$ statements of size
bounded by $2\cdot |s|$. In fact, $2^{|S|}\cdot |s|^{2|S|}$
is also a bound for the total number of different conjunctions
that may appear along the inference process. Thus, if we
treat each conjunction once, by removing the generated ones
that have been already treated, at most $2^{|S|}\cdot |s|^{2|S|}$
inference steps are executed. Each inference step takes time
proportional to the size of a conjunction, which is
bounded by $|S|\cdot 2\cdot |s|$, multiplied by the
maximum arity, which is $2$ by our simplifying assumption.
Thus, the total cost is
$\OO(2^{|S|}\cdot |s|^{2|S|+1}|S|)$.
\end{proof}

\subsection{The algorithm}

We summarize in Figure~\ref{alg-one} the EXPTIME algorithm for deciding
regularity of a set of terms with regular constraints.

\begin{figure*}
\begin{small}
\begin{tabbing}
oo \= oo \= oo \= oo \= oo \= oo \kill
Input: set of terms with regular constraints $\langle S_1, M_1\rangle$ \\
Compute set of terms with single regular constraints $\langle S_2, R_1\rangle$. \\
$W:=\emptyset$.\\
For each non-regular $s\in S_2$ do:\\
\>    Determine $s$ at $\Posnv(S_2)$,
giving $\{s_1,...,s_k\}$.\\ 
\>    For each $i=1$ to $k$ do:\\
\>\>        Compute $S_3:=\{t\in S_2\mid t$ structurally subsumes $s_i\} - \{s_i\}$.\\
\>\>        Run \emph{infinite-instances}$(s_i,S_3,A,C,W,|Q|+2H)$:\\
\>\>\>            $\{$ Build formula $\mathcal{F}(s_i,S_3,W,|Q|+2H)$, \\
\>\>\>             $\rho = \text{Reduce}(\mathcal{F})$, \\
\>\>\>             Return $(\rho \neq \text{empty formula})$. $\}$ \\
\>\>        If infinite-instances returns true, then Return(``not-regular'');\\
\>    $W:=W\cup\{x\mid x\text{ occurs }\geq 2\text{ times in }s\}$.\\
Return(``regular'');
\end{tabbing}
\end{small}
\caption{The EXPTIME algorithm for deciding regularity.}
\label{alg-one}
\end{figure*}

The algorithm starts by transforming the input instance $\langle
S_1,M_1\rangle$ into an equivalent set of terms with single regular
constraints in exponential time, according to
Lemma~\ref{lemma-transformationtosingle}. Then, the algorithm
(implicitly) considers a restricted regular constraint $R_2=\langle A,
V, C, W, |Q|+2H\rangle$, where $W=\emptyset$ at the beginning. The
determination of $s$ into $s_1,\ldots,s_k$ is done according to
Lemma~\ref{lemma-determining}.  This determination also takes
exponential time on the size of the input instance $\langle S_1,
M_1\rangle$, since $\Prefixes(\Posnv(S_2))$ coincides with
$\Prefixes(\Posnv(S_1))$. Finally, the infinite-instances property can
be also determined in exponential time, due to Lemma~
\ref{lem:infiniteinstancesalgorithm} and the fact that
 $|S_3|\leq |S_1|-1$. Thus, it follows from the previous lemmas that
the algorithm runs in exponential time with respect to
$\|S_1\|+|M_1|$.

Now we discuss the correctness of the algorithm. Let $R_2'$ be
the extension of $R_2$ obtained when determining $s$ into
$s_1,\ldots,s_k$, according to Lemma~\ref{lemma-determining}.
Assume the case where none of the $s_i$ satisfies the infinite
instances property in $S_2$, and consider a concrete term $s_i$
satisfying that $\langle \{s_i\},R_2\rangle$ has instances not in
$\LL(\langle S_2-\{s\},R_2\rangle)$.  By
Lemma~\ref{lem:infiniteinstancesalgorithm}, $s_i$
cannot have duplicated variables with associated infinite
language. Now, consider a duplicated variable $x$ with infinite
language and occurring at a position $p$ in $s$.
By Lemma~\ref{lemma-determining}, $s_i/p$
has height bounded by $H$,
and occurs at another position in $s_i$. Thus, all
the variables $y$ occurring in $s_i/p$ are duplicated
in $s_i$, and hence, $\LL(\langle\{y\},R_2'\rangle)$ is finite.
In particular, they can be instantiated by a term with
height bounded by $|Q|$ in order to get an instance.
Therefore, $\LL(\langle\{s_i/p\},R_2'\rangle)$ is finite,
and any instance $t$ of $\LL(\langle\{s_i\},R_2'\rangle)$
satisfies $\height(t/p)\leq |Q|+H$.

{From} the above considerations we conclude that
all instances $t$ of $\langle \{s\},R_2\rangle$
not in $\LL\langle S-\{s\},R_2\rangle$ satisfy
the following statement: for each position $p$ with a duplicated
variable in $s$, $\height(t/p)\leq |Q|+H\leq|Q|+2H$.
Thus, by adding the duplicated variables in $s$ to $W$
we preserve the represented language.

\begin{theorem}\label{theo-regins}
The above algorithm solves RITRC in exponential time.
\end{theorem}

\section{Concluding Remarks}

In this contribution we have shown that the RITRC problem is EXPTIME-hard, and
have presented a new algorithm that solves RITRC
in exponential time. This problem is a particular
case of the HOM problem~\cite{ful94}: given a DTA $A$ and a
tree homomorphism $H$, is $H(\LL(A))$ regular?
The decidability of this problem is a long-standing open question.
The main problem is how to handle non-linearity of $H$, 
and to determine in which cases it forces non-regularity of $H(\LL(A))$.
Our algorithm gives some intuition about when
non-linearity poses a real problem for the
regularity of the represented set (it also gives
an exponential time solution for the HOM problem 
in the case that non-linear rules
are only applied at bounded depth of the input tree,
cf.~\cite{godmantis08}).
But, it is still far from solving the general problem.

\comment{
A (tree) homomorphism $h$ is given by a mapping that associates 
with every $f\in\Sigma^{(k)}$ a term in $T_\Delta$. Its
translation $\hat{h}:T_\Sigma\to T_\Delta$ is defined by 
interpreting $h$ as rewrite rules in the obvious way. 
Let $L\subseteq T_\Sigma$ be regular (given by a DTA $A$).
Is is a long-standing open problem whether regularity 
is decidable for $\hat{h}(L)$.
It follows from the results of~\cite{godmantis08} that 
if $h$ is \emph{top-copying} for $L$, i.e., every $h(\sigma??)$ with
duplicated variables is only applied at bounded depth
of any input tree in $L$, then regularity of $\hat{h}(L)$ can be
reduced (in polynomial time) to testing regularity of a
set of terms with regular constraints.

\begin{corollary}
Regularity of $\hat{h}(\LL(A))$ is decidable in exponential time
if  $h$ is a top-copying homomorphism for $\LL(A)$.
\end{corollary}
}

\bibliographystyle{siam}
\bibliography{bib}

\begin{thebibliography}{10}

\bibitem{bencashospievan08}
{\sc V.~Benzaken, G.~Castagna, H.~Hosoya, B.C. Pierce, and S.~Vansummeren},
  {\em The Encyclopedia of Database Systems}, Springer, 2009, ch.~``{XML}
  Typechecking''.

\bibitem{tata2007}
{\sc H.~Comon, M.~Dauchet, R.~Gilleron, C.~L\"oding, F.~Jacquemard, D.~Lugiez,
  S.~Tison, and M.~Tommasi}, {\em Tree automata techniques and applications}.
\newblock Available at http://www.grappa.univ-lille3.fr/tata, 2007.

\bibitem{ComonDelor94}
{\sc H.~Comon and C.~Delor}, {\em Equational formulae with membership
  constraints}, Infor. and Comput., 112 (1994), pp.~167--216.

\bibitem{don70}
{\sc J.~Doner}, {\em Tree acceptors and some of their applications}, J. Comp.
  Syst. Sci., 4 (1970), pp.~406--451.

\bibitem{ful94}
{\sc Z.~F{\"u}l{\"o}p}, {\em Undecidable properties of deterministic top-down
  tree transducers}, Theoret. Comput. Sci., 134 (1994), pp.~311--328.

\bibitem{gecste97}
{\sc F.~G\'ecseg and M.~Steinby}, {\em Tree languages}, in Handbook of Formal
  Languages, Volume 3, G.~Rozenberg and A.~Salomaa, eds., Springer, 1997,
  ch.~1.

\bibitem{godmantis08}
{\sc G.~Godoy, S.~Maneth, and S.~Tison}, {\em Classes of tree homomorphisms
  with decidable preservation of regularity}, in FoSSaCS, vol.~4962 of LNCS,
  Springer, 2008, pp.~127--141.

\bibitem{DBLP:journals/jfp/HosoyaP03}
{\sc H.~Hosoya and B.~C. Pierce}, {\em Regular expression pattern matching for
  {XML}}, J. Funct. Program., 13 (2003), pp.~961--1004.

\bibitem{DBLP:journals/toplas/HosoyaVP05}
{\sc H.~Hosoya, J.~Vouillon, and B.~C. Pierce}, {\em Regular expression types
  for {XML}}, ACM Trans. Program. Lang. Syst., 27 (2005), pp.~46--90.

\bibitem{kucrus99}
{\sc G.~Kucherov and M.~Rusinowitch}, {\em Patterns in words versus patterns in
  trees: A brief survey and new results}, in Ershov Memorial Conference,
  Springer, 1999, pp.~283--296.

\bibitem{lasmar87}
{\sc J.-L. Lassez and K.~Marriott}, {\em Explicit representation of terms
  defined by counter examples}, J. Automat. Reason., 3 (1987), pp.~301--317.

\bibitem{mezwri67}
{\sc J.~Mezei and J.~B. Wright}, {\em Algebraic automata and context-free
  sets}, Inform. and Control, 11 (1967), pp.~3--29.

\bibitem{DBLP:journals/toit/MurataLMK05}
{\sc M.~Murata, D.~Lee, M.~Mani, and K.~Kawaguchi}, {\em Taxonomy of {XML}
  schema languages using formal language theory.}, ACM Trans. Internet Techn.,
  5 (2005), pp.~660--704.

\end{thebibliography}






\end{document}